\documentstyle[11pt,epsf]{article}
\newcommand{\simgt}{\lower.5ex\hbox{$\; \buildrel > \over \sim \;$}}
\newcommand{\simlt}{\lower.5ex\hbox{$\; \buildrel < \over \sim \;$}}
\begin{document}

\begin{minipage}[c]{8cm}
RESCEU-2/99\\
UTAP-317/98\\
Advances in Space Research, in press.
\end{minipage}\\

{\Large \bf COSMOLOGICAL IMPLICATIONS OF GALAXY CLUSTERS \\
IN X-RAY, MILLIMETER, AND SUBMILLIMETER BANDS}

Yasushi Suto$^{1,2}$, Tetsu Kitayama$^2$, Eiichiro Komatsu$^3$, Makoto
Hattori$^3$, Ryohei Kawabe$^4$, Hiroshi Matsuo$^4$, 
Sabine Schindler$^5$, and Kohji Yoshikawa$^6$

\vspace{0.5cm}
$^1${\it Research Center for the Early Universe, School of Science, 
University of Tokyo, Tokyo 113-0033, Japan }\\
$^2${\it Department of Physics, University of Tokyo, 
         Tokyo 113-0033, Japan }\\
$^3${\it Astronomical Institute, T\^{o}hoku University, 
         Sendai 980-8578, Japan }\\
$^4${\it Nobeyama Radio Observatory, 
         Nagano 384-1305, Japan }\\
$^5${\it Astrophysics Research Institute, Liverpool John Moores
         University, Liverpool L3 3AF, UK }\\ 
$^6${\it  Department of Astronomy, Kyoto University, 
          Kyoto 606-8502, Japan }

\vspace{0.5cm}

\section*{ABSTRACT}

Cosmological implications of clusters of galaxies are discussed with
particular attention to their importance in probing the cosmological
parameters. More specifically we compute the number counts of clusters
of galaxies, Log $N$ -- Log $S$ relation, in X-ray and submm bands on
the basis of the Press--Schechter theory. As an important step toward
breaking the degeneracy among the viable cosmological models, we
observed the most luminous X-ray cluster RXJ1347-1145 in three bands
(21 and 43 GHz in the Nobeyama Radio Observatory, Japan, and 350 GHz
in the J. C. Maxwell telescope at Mauna Kea, Hawaii).  We report on
the preliminary results which are in good agreement with the profile
of the Sunayev -- Zel'dovich effect predicted on the basis of the
previous X-ray observation of the cluster.

\section{INTRODUCTION}

There are several reasons why clusters of galaxies are regarded as
useful tools in cosmology; (i) since the dynamical time-scale of
clusters is comparable to the age of the universe, they should retain
the cosmological initial conditions fairly faithfully, (ii) clusters
can be observed in various wavelengths including optical, X-ray,
radio, mm and submm bands, and recent and future big projects (e.g.,
SDSS, AXAF, PLANCK) aim to make extensive surveys and detailed
imaging/spectroscopic observations of clusters, (iii) to the first
order approximation, clusters can be regarded as a simple system of
dark matter, gas and galaxies, and thus theoretically well-defined and
relatively well-understood, at least compared with galaxies
themselves, (iv) clusters of galaxies can be observed up to high
redshifts and thus provide a probe of the distant universe.

X-ray observations are particularly suited for the study of clusters,
because the X-ray emissivity is proportional to $n_e^2$ and less
sensitive to the projection contamination which has been known to be a
serious problem in their identifications with the optical data. In
addition, the Sunyaev -- Zel'dovich (SZ) effect (Sunyaev \& Zel'dovich
1972), observed in radio, millimeter and submillimeter bands, is now
opening a new window to investigating cluster properties, especially
at high redshifts.  In this paper, we aim to show the significance of
current and future observations in these bands in cosmology.

\section{LOG $N$ -- LOG $S$ OF X-RAY CLUSTERS}

We compute the number of clusters observed per unit
solid angle with X-ray flux greater than $S$ by
\begin{eqnarray}
  N(>S)= \int_{0}^{\infty}dz ~d_A^2(z) \, c
  \left|{\frac{dt}{dz}}\right| \int_{S}^\infty dS ~ (1+z)^3 n_M(M,z)
  \frac{dM}{dT_{gas}}\frac{dT_{gas}}{dL_{band}} \frac{dL_{band}}{dS},
\label{eq:logns}
\end{eqnarray}
where $c$ is the speed of light, $t$ is the cosmic time, $d_A$ is the
angular diameter distance, $T_{gas}$ and $L_{band}$ are respectively
the gas temperature and the band-limited absolute luminosity of
clusters, and $n_M(M,z)dM$ is the comoving number density of
virialized clusters of mass $M \sim M+dM$ at redshift $z$.

Given the observed flux $S$ in an X-ray energy band [$E_a$,$E_b$], the
source luminosity $L_{band}$ at $z$ in the corresponding band
[$E_a(1+z)$,$E_b(1+z)$] is written as
\begin{equation}
  L_{band}[E_a(1+z),E_b(1+z)] = 4 \pi d_{L}^2(z) S[E_a,E_b],
\label{eq:ls}  
\end{equation}
where $d_{L} = (1+z)^2 d_A$ is the luminosity distance. We adopt the
observed $L_{bol} - T_{gas}$ relation parameterized by
\begin{equation}
  L_{bol} = L_{44} \left( \frac{T_{gas}}{6{\rm keV}} 
\right)^{\alpha}
  (1+z)^\zeta ~~ 10^{44} h^{-2}{\rm ~ erg~sec}^{-1} .
\label{eq:lt}
\end{equation}
We take $L_{44}=2.9$, $\alpha=3.4$ and $\zeta=0$ as a fiducial set of
parameters on the basis of recent observational indications (David et
al. 1993; Mushotzky \& Scharf 1997).  Then we translate
$L_{bol}(T_{gas})$ into the band-limited luminosity
$L_{band}[T_{gas},E_1,E_2]$ by properly taking account of metal line
emissions in addition to the thermal bremsstrahlung (we fix the
abundance of intracluster gas as 0.3 times the solar value).

Assuming that the intracluster gas is isothermal, its temperature
$T_{gas}$ is related to the total mass $M$ by
\begin{eqnarray}
  k_{B} T_{gas} &=& \gamma {\mu m_p G M \over 3 r_{\rm vir}(M,z_f)},
  \nonumber \\ &=& 5.2\gamma (1+z_f) \left({\Delta_{\rm vir} \over
      18\pi^2}\right)^{1/3} \left({M \over 10^{15} h^{-1} M_\odot}
  \right)^{2/3} \Omega_0^{1/3} ~{\rm keV}.
\label{eq:tm}
\end{eqnarray}
where $k_B$ is the Boltzmann constant, $G$ is the gravitational
constant, $m_p$ is the proton mass, $\mu$ is the mean molecular weight
(we adopt $\mu=0.59$), and $\gamma$ is a fudge factor of order unity
which may be calibrated from hydrodynamical simulations or
observations.  We adopt $\gamma=1.2$ as canonical based on the recent
simulation results (e.g. Bryan \& Norman 1998).  The virial radius
$r_{\rm vir}(M,z_f)$ of a cluster of mass $M$ virialized at $z_f$ is
computed from $\Delta_{\rm vir}$, the ratio of the mean cluster
density to the mean density of the universe at that epoch. We evaluate
this quantity using the formulae for the spherical collapse model
presented in Kitayama \& Suto (1996b) and assuming for simplicity that
$z_f$ is equal to the epoch $z$ at which the cluster is observed.
Finally, we compute the mass function $n_M(M,z)dM$ in equation
(\ref{eq:logns}) using the Press--Schechter theory (Press \& Schechter
1974) assuming $z=z_f$ as above.  The effect of $z_f \neq z$ is
discussed by Kitayama \& Suto (1997) in this context, and the more
general consideration of $z_f \neq z$ is given in Lacey \& Cole
(1993), Sasaki (1994), and Kitayama \& Suto (1996a,b).

\section{BREAKING THE DEGENERACY \\ WITH THE SUNYAEV
-- ZEL'DOVICH EFFECT }

\begin{figure}[h]
\begin{center}
\leavevmode\epsfysize=8.5cm \epsfbox{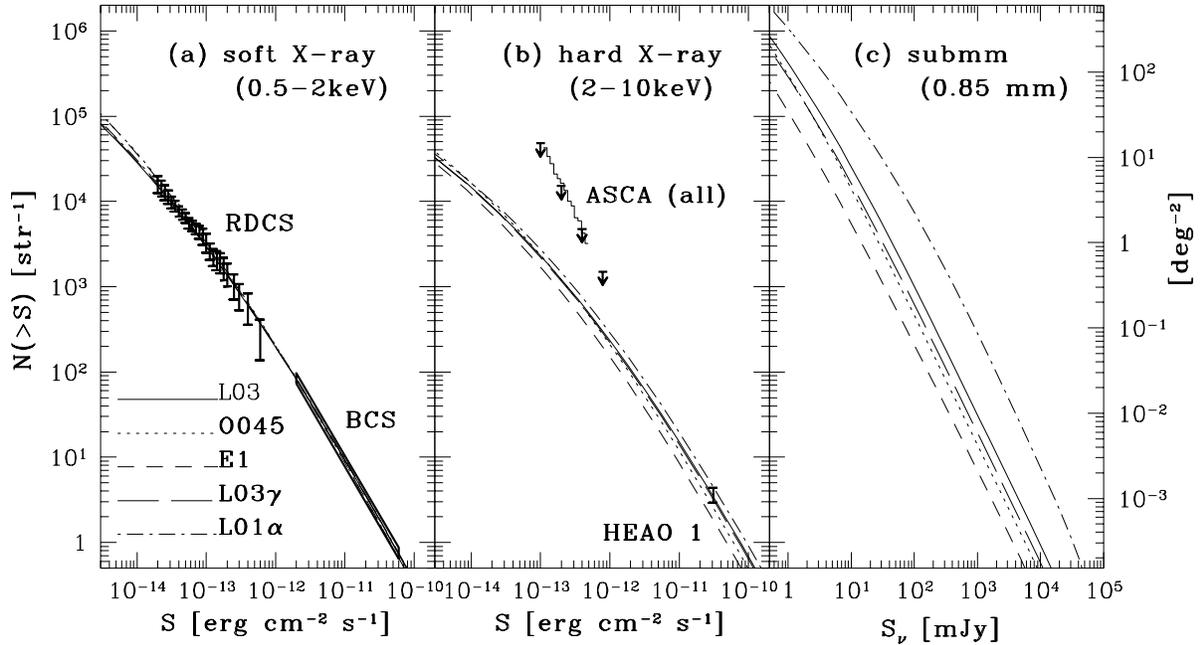}
\caption{
  Log$N$--log$S$ relations of galaxy clusters for CDM models in (a)
  the soft X-ray (0.5--2 keV) band, (b) the hard X-ray (2--10 keV)
  band, and (c) the submm (0.85 mm) band. Lines represent the models
  listed in Table 1; L03 (solid), O045 (dotted), E1 (short dashed),
  L03$\gamma$ (long dashed), and L01$\alpha$ (dot-dashed).  Also shown
  in panel (a) are the 1$\sigma$ error bars from the ROSAT Deep
  Cluster Survey (Rosati et al. 1995, 1997), and the error box from
  the ROSAT Brightest Cluster Sample (Ebeling et al. 1997, 1998).  In
  panel (b), the number of clusters observed by {\it HEAO 1}
  (Piccinotti et al. 1982) is indicated by the 1$\sigma$ error bar,
  and the counts of {\it all} X-ray sources detected by {\it ASCA} are
  plotted for reference as a histogram (Ueda et al. 1998) and downward
  arrows (Cagnoni et al.  1998).}
\label{fig:ns3} 
\end{center}
\end{figure}

As has been realized for a while, a cosmological observation is often
accounted for by a set of quite different models.  In the present
case, we found that the following models equally fit the Log $N$ --
Log $S$ of ROSAT X-ray clusters (quoted numbers are 1$\sigma$
statistical and systematic errors, respectively):
\begin{equation}
\label{eq:sigma8logns}
\sigma_8 = (0.54 \pm 0.02 \pm 0.08) \times
\Omega_0^{-0.35-0.80\Omega_0+0.55\Omega_0^2}
\end{equation}
for $\lambda_0=1-\Omega_0$ Cold Dark Matter (CDM) models, and
\begin{equation}
\sigma_8 = (0.54 \pm 0.02 \pm 0.08) \times
        \Omega_0^{-0.28-0.91\Omega_0+0.68\Omega_0^2} 
\end{equation}
for $\lambda_0=0$ CDM models.  In order to exhibit this {\it
  degeneracy} among viable cosmological models, we select several
examples listed in Table 1 which reproduce the ROSAT Log $N$ -- Log
$S$ in almost an indistinguishable manner (Fig.\ref{fig:ns3}a).

\begin{table}[h]
\caption{CDM models from the ROSAT X-ray
  Log $N$ -- Log $S$.}
\begin{center}
\begin{tabular}{ccccccc}
\hline\\[-10pt]
Model & $\Omega_0$ &  $\lambda_0$  
&  $h$ &   $\sigma_8$ & $\alpha$ & $\gamma$\\ 
[4pt]\hline \\[-6pt]
L03 & 0.3  & 0.7 & 0.7 & 1.04 & 3.4 & 1.2  \\
O045 & 0.45 & 0   & 0.7 & 0.83 & 3.4 & 1.2  \\
E1 & 1.0  & 0   & 0.5 & 0.56 & 3.4 & 1.2 \\
L03$\gamma$ & 0.3  & 0.7 & 0.7 & 0.90 & 3.4 & 1.5 \\
L01$\alpha$ & 0.1  & 0.9 & 0.7 & 1.47 & 2.7 & 1.2 \\
\hline
\end{tabular}
\end{center}
\end{table}

While this degeneracy can be broken by several methods, one
possibility is to combine with observations of the Sunyaev \&
Zel'dovich (SZ) effect, the distortion of the cosmic microwave
background (CMB) spectrum due to the inverse-Compton scattering in
high temperature intracluster electrons (Sunyaev \& Zel'dovich 1972,
SZ).  The specific intensity of the SZ signal at the rest frame of the
cluster is given by
\begin{eqnarray} 
I_\nu^{\rm NR} &=& \frac{2 h_{\rm P} \nu^3}{c^2} \tilde{n} =   
i_0 (1+z)^3 {x^3 \over e^x-1} +  \Delta I_{\nu}^{\rm NR}, \nonumber \\
  \Delta I_{\nu}^{\rm NR} &=& i_0 (1+z)^3 g(x) y, 
\label{eq:szi}
\end{eqnarray}
where $h_{\rm P}$ is the Planck constant, and $y$ is the Compton
$y$-parameter:
\begin{equation}
y \equiv \int \frac{k_{\rm B}T}{m_{\rm e}c^2} 
\sigma_{\rm T} n_{\rm e} c d  t.    
\end{equation} 
In the above expressions, $T_{2.73}$ is the CMB temperature $T_{\gamma
  0}$ divided by $2.73$K, $x$ is the photon energy in units of $k
T_{\gamma 0}$, and
\begin{eqnarray} 
i_0 &=& \frac{2(k_{\rm B} T_{\gamma 0})^3}{(h_{\rm P}c)^2} = 2.29
\times 10^4 T_{2.73}^3 \mbox{~~mJy arcmin$^{-2}$}, \\
g(x) &=& \frac{x^4e^x}{(e^x-1)^2} 
     \left(x\coth\frac{x}{2}-4 \right) .
\label{eq:gx}
\end{eqnarray} 
By integrating equation (\ref{eq:szi}) over a cluster, one obtains the
excess flux from the cluster relative to the CMB:
\begin{eqnarray}
S_\nu^{\rm NR}  &=& \frac{1}{d_{\rm L}^2(z)}\int \Delta
I_{\nu}^{\rm NR} d  A,  
\nonumber   \\ 
&=& 4.9 g(x) {2-Y \over 2} 
\left[\frac{d_{\rm A}(z)}{c H_0^{-1}}\right]^{-2}
\left({M \over 10^{15} h^{-1} M_\odot} \right)
\left(\frac{T}{\rm keV}\right)
\frac{\Omega_{\rm B}h}{\Omega_0} ~{\rm mJy} . 
\label{eq:szflux}
\end{eqnarray}

\begin{figure}[tbhp]
\begin{center}
\leavevmode\epsfysize=8cm \epsfbox{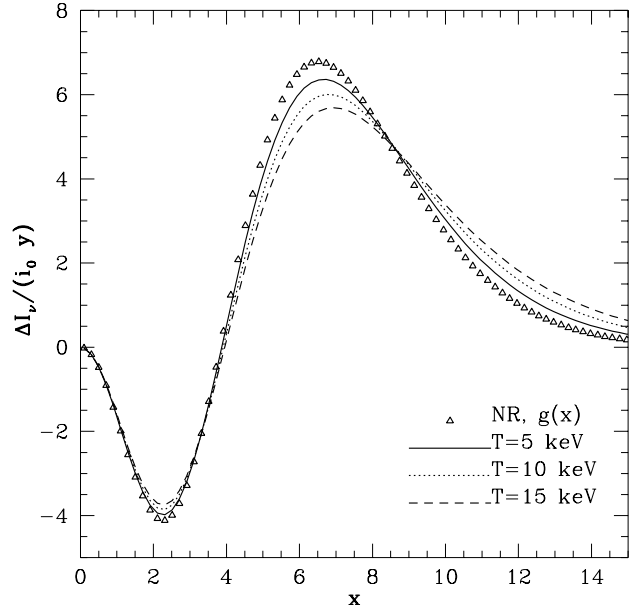}
\caption{
 The spectral dependence of the intensity change
  $\Delta I_\nu/(i_0 y)$ induced by the SZ effect. The
  non-relativistic expression $g(x)$ is shown as triangles, whereas
  the relativistic results are plotted as lines for $T=5$ keV (solid),
  $10$ keV (dotted), and $15$ keV (dashed).}
\label{fig:szrel}
\end{center}
\end{figure}

The spectrum of the SZ effect is completely specified by the function
$g(x)$ which is independent of cluster properties. Most importantly,
$g(x)$ is negative at $x<3.83$ ($\nu <217$GHz, $\lambda > 0.14$cm),
while positive at $x> 3.83$ (Fig.\ref{fig:szrel}).  Clusters of
galaxies therefore serve as negative (absorbing) sources at mm and cm
bands, and as positive (emitting) sources at submm band.  At $T \simgt
10$ keV, relativistic effects of electrons become important and the
above expressions need to be modified. We thus apply the relativistic
correction to the intensity change derived by Rephaeli (1995; see also
Rephaeli \& Yankovitch 1997): At the observed wavelength of 0.85 mm,
the above correction leads to 4\%, 11\% and 16\% reduction in the
excess flux, i.e., $g(x)$, at $T =3$ keV, $8$ keV and $12$ keV,
respectively. While clusters with $T \simgt 10$ keV are rare and the
above correction does not make significant difference, we take it into
account for completeness.

\begin{figure}[h]
\begin{center}
\leavevmode\epsfxsize=14cm \epsfbox{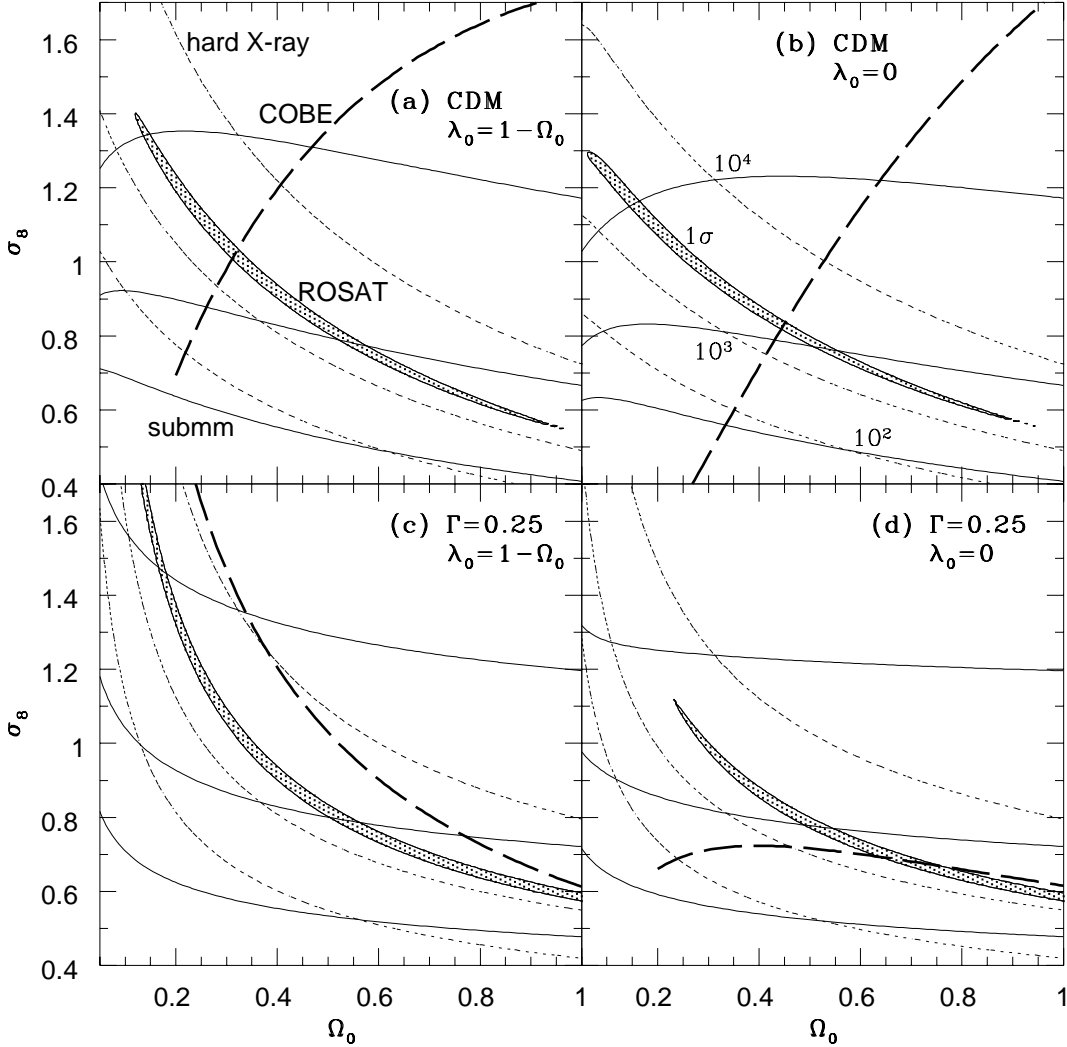}
\caption{
  Contour maps on the $\Omega_0$-$\sigma_8$ plane in (a) spatially
  flat ($\lambda_0=1-\Omega_0$) CDM models, (b) open ($\lambda_0=0$)
  CDM models, (c) spatially flat CDM-like models with the fixed shape
  parameter ($\Gamma=0.25$), and (d) open CDM-like models with
  $\Gamma=0.25$.  In all cases, $h=0.7$ and the fiducial set of
  intracluster gas parameters is assumed. Shaded regions represent the
  1$\sigma$ significance contours from the soft X-ray (0.5--2 keV)
  log$N$--log$S$. Dotted and solid lines indicate the contours of the
  number of clusters greater than $S$ per steradian ($10^2$, $10^3$,
  $10^4$ from bottom to top) with $S = 10^{-13}$
  erg\,cm$^{-2}$\,s$^{-1}$ in the hard X-ray (2--10 keV) band and with
  $S_\nu = 50$ mJy in the submm (0.85 mm) band, respectively.  Thick
  dashed lines represent the {\it COBE} 4 year results by Bunn \&
  White (1997).}
\label{fig:cont}
\end{center}
\end{figure}

Panels (b) and (c) in Figure \ref{fig:ns3} show the number counts of
clusters of galaxies in the hard X-ray (2--10 keV) and the submm (0.85
mm, due to the SZ effect) bands for models which reproduce the
observed Log $N$ -- Log $S$ in the soft X-ray band (Table 1). Clearly
the predictions in submm bands are significantly different among the
models.  To see this clearly, we plot in Figure \ref{fig:cont} contour
maps of the cluster log$N$--log$S$ in different bands on the
$\Omega_0-\sigma_8$ plane. For the 0.5--2 keV band, the 1$\sigma$
significance level is plotted, while for the 2--10 keV and submm
bands, the contours of the number of clusters per steradian ($10^2$,
$10^3$, $10^4$ from bottom to top) at $S(\mbox{2--10 keV})=10^{-13}$
erg\,cm$^{-2}$\,s$^{-1}$ and $S_\nu(\mbox{0.85 mm})=50$ mJy are
plotted, respectively. The $\sigma_8$ values derived from the {\it
  COBE} 4 year data (Bunn \& White 1997) are also given for reference.
Panels (a) and (b) show that the contours for the 2--10 keV band
counts run almost parallel to the $\chi^2$ contour of the 0.5--2 keV
band counts. In this sense, future log$N$--log$S$ data in the hard
X-ray band will provide an independent consistency check of the {\it
  ROSAT} soft X-ray data.  The shape of the submm log$N$--log$S$
contours, on the other hand, is quite different, especially at high
$\sigma_8$, and thus should place complementary constraints on
$\Omega_0$ and $\sigma_8$.  If the fixed shape parameter
($\Gamma=0.25$) is adopted, the corresponding log$N$--log$S$ contours
are similar to the conventional CDM ($\Gamma\simeq \Omega_0 h$) case,
except at $\Omega_0 \simlt 0.2$ and $\Omega_0 \simgt 0.8$.  By
contrast, the {\it COBE} normalized $\sigma_8$ is very sensitive to
any changes in the spectral shape.

\section{SEARCH FOR THE SUBMM SZ EFFECT OF RXJ 1347-1195}

RXJ 1347.5-1145 ($z=0.45$) is the most luminous X-ray cluster of
galaxies known so far (Schindler et al. 1997).  Its exceptionally high
column density of electrons at the center implies that this is one of
the most promising clusters for the detection of the first unambiguous
SZ increment signal. This is clearly illustrated in Table 2 where we
compute the expected central SZ intensity $\Delta I_\nu(0)$ at 0.85mm
for several clusters with measured SZ temperature decrement in the
Rayleigh -- Jeans regime (except for our current target, RXJ
1347-1195).  Wherever available, 1$\sigma$ errors in the parameters
are shown which are also adopted to estimate the error in $\Delta
I_\nu(0)$.

\vspace*{1cm}
\begin{table}[h]
\vspace*{-0.5cm}
\caption{ Expected central SZ intensity $\Delta I_\nu(0)$ at
0.85mm for several clusters}
\begin{center}
{\footnotesize 
\begin{tabular}{c|c|c|c|c|c|c}
cluster & $z$ & $T$ [keV] & $\beta$ & $\theta_c$ [arcsec] &
  $n_{\rm e}^0 [10^{-2}h_{50}^{1/2}\mbox{cm}^{-3}]$ & 
  $\Delta I_\nu(0) [h_{50}^{-1/2}\mbox{mJy/beam}]$\\ \hline  
RXJ 1347--1145 & 0.45 & $9.3^{+1.1}_{-1.0}$ & $0.56\pm{0.04}$ & $8.4\pm{1.8}$& 
9.4 & $7.1 \pm 1.7$\\
\hline
CL0016+16 & 0.555 & $8.4^{+1.2}_{-0.6}$ & 0.81 & 51 & 0.65 & 
$1.9^{+0.3}_{-0.1}$   \\ 
A2163 & 0.201 & $14.6 \pm 0.5$ & $0.62 \pm 0.01$ & $72 \pm 3$ & $0.77
  \pm 0.03 $ & $4.5 \pm 0.3$  \\
A2218 & 0.171 & $6.7 \pm 0.45$ & $0.65^{+0.05}_{-0.03}$ & $60 \pm
  12$ &  0.63 & $1.2 \pm 0.3$ \\
A2142 & 0.0899  & $8.68 \pm 0.12$ & $1.0\pm 0.3$ &$221\pm 84 $& $0.493
  \pm 0.023$  &  $1.6 \pm 0.8$\\
A478 & 0.0881 & $6.56\pm 0.09$ & $0.67 \pm 0.03$ &  $116\pm 18$ &
  $0.675\pm 0.124$ & $1.3 \pm 0.3 $ \\
A2256 & 0.0581 & $7.51 \pm 0.11$ & $0.80 \pm 0.02$ & $320 \pm 12$ &
  $0.251 \pm 0.013$ & $0.88 \pm 0.06$ \\
Coma & 0.0235 & $9.10 \pm 0.40$ & $0.75 \pm 0.03$  & $630 \pm 36$ &
  $0.289 \pm 0.004$ &  $1.1 \pm 0.1$\\
\end{tabular}}
\end{center}
\end{table}

On the basis of this analysis, we performed the search for the SZ
effect towards the cluster RXJ1347-1145, at 21 and 43 GHz with NRO
(Nobeyama Radio Observatory) 45m telescope between March 3 and 15,
1998, and at 350 GHz with SCUBA installed on JCMT (J.C.Maxwell
Telescope) 15m telescope on May 30 and 31, 1998. Although there have
been a few claims of the detection of the SZ temperature increment for
some clusters (e.g. Andreani et al. 1996; Lamarre et al. 1998), they
measured the total flux alone which could be significantly
contaminated by the dust in our Galaxy and also by submm sources in
the field. Therefore it is essential to map the profile of a cluster
so that one can test it against the predicted SZ signal profile. This
is the reason why we attempted the observation with SCUBA which has a
reasonable angular resolution, $15''$ in FWHM, with a FOV of $160''$.

As a matter of fact, it turned out that a radio point source exists
near the center of our target cluster. The flux of the central point
source had been determined to be $47.6\pm 0.45$ mJy at 1.4GHz with VLA
(Condon et al. 1988), and $11.0\pm 0.7$ mJy at 28.5GHz in the OVRO
serendipitous survey of SZ effect (J.Carlstrom, private
communication). We also observed the central source at 100 GHz with
Nobeyama Millimeter Array (NMA) between May 19th to 21st, 1998, and
found the flux to be $5.0\pm1.5$ mJy.  The three data at $\nu\le 100$
GHz are very well approximated by a single power-law:
\begin{equation}
  \label{eq:pointflux}
  F_{\rm p}(\nu)=(56.1\pm 1.0)(\nu/1{\rm GHz})^{-0.49\pm 0.02} \, {\rm mJy} ,
\end{equation}
where the quoted errors are $1\sigma$.  We subtracted the contribution
of the point source according to equation (\ref{eq:pointflux}) from
our data at 21 and 43 GHz.  Although the point source might be a
variable, the degree of variability is known to be fairly small for
radio sources with steep spectrum (spectral index $\simlt -0.5$;
e.g. Eckart, Hummel \& Witzel 1989).

\begin{figure}[tbhp]
\begin{center}
  \leavevmode\epsfxsize=17cm \epsfbox{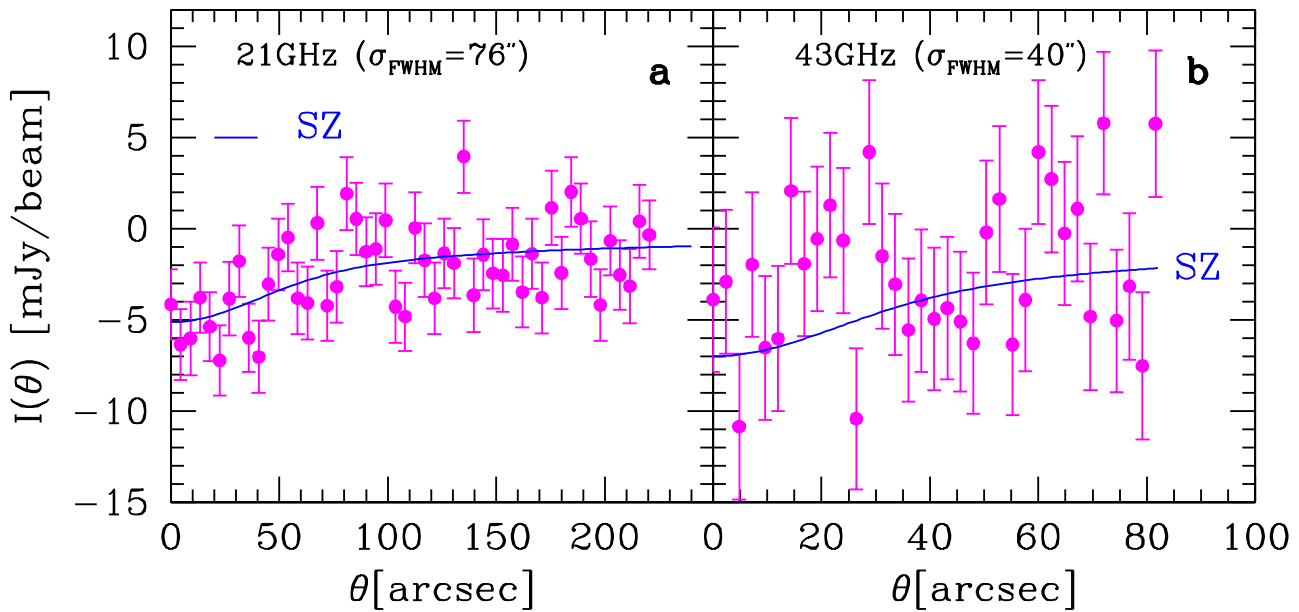}
\caption{\baselineskip=14pt
  Radial intensity profile towards RXJ1347 at (a) 21 and (b) 43 GHz
  observed at NRO. The estimated contribution of the central radio
  source (12.7mJy at 21 GHz, and 8.9 mJy at 43 GHz) is subtracted.
  Filled circles indicate our data with $1\sigma$ error-bars. Solid
  lines indicate the prediction of the SZ signal (without the point
  source contribution) using the best-fit parameters in the X-ray
  observation.}
\label{fig:2143}
\end{center}
\end{figure}

Figure \ref{fig:2143} shows the profiles of the scans towards the
cluster at 21 and 43 GHz after subtracting the point source.  In both
cases, a extended negative intensity profile is clearly visible.
Especially at 21 GHz where the S/N is significantly higher than at
43GHz, the profile is quite consistent with the SZ profile expected
from the X-ray observation. This is the first detection of the SZ
temperature decrement towards the cluster RXJ 1347-1145.

\begin{figure}[tbhp]
\begin{center}
  \leavevmode\epsfysize=8cm \epsfbox{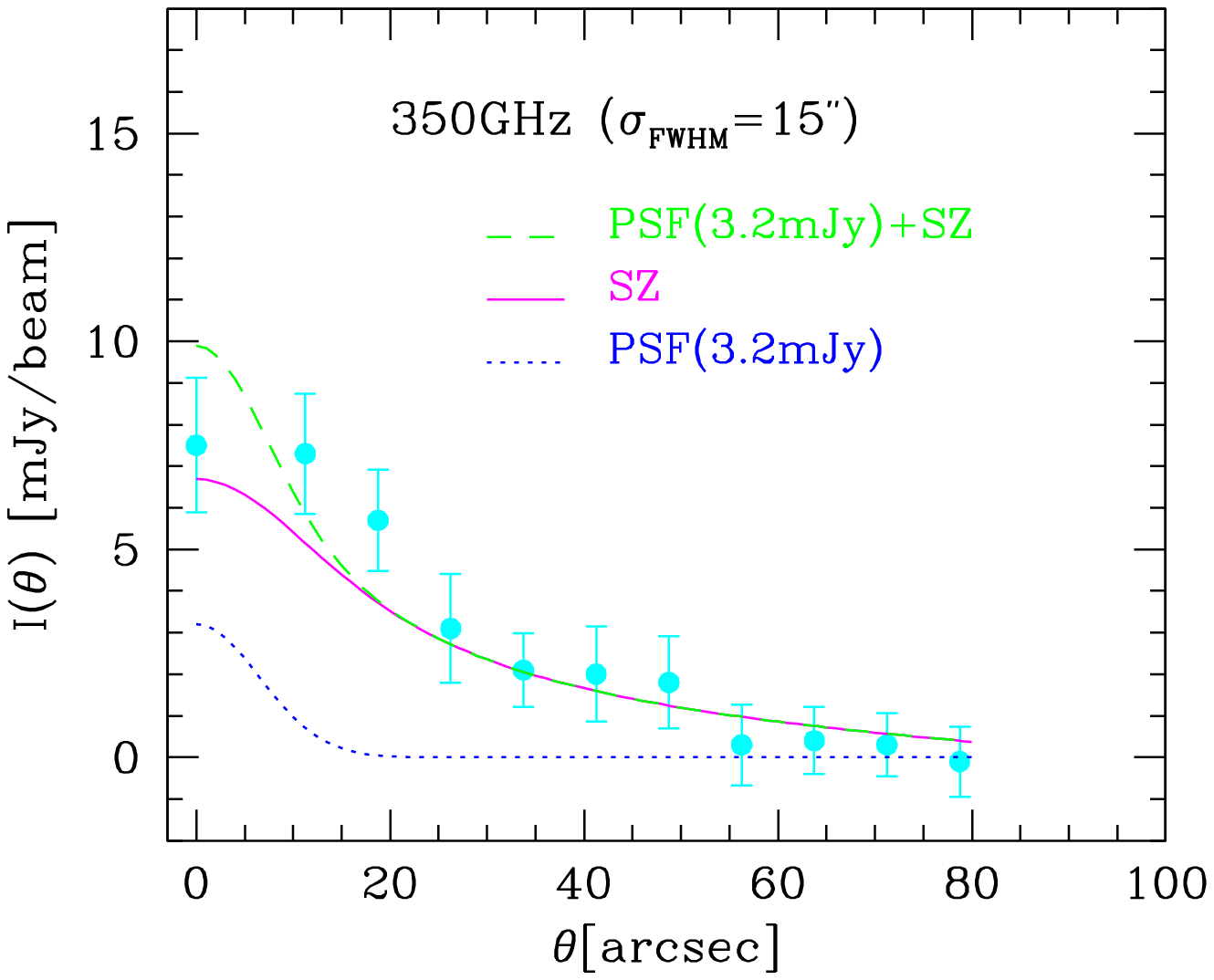}
\caption{\baselineskip=14pt
  Radial intensity profile towards RXJ1347 at 350 GHz observed at
  JCMT/SCUBA.  Filled circles indicate our data with $1\sigma$
  error-bars. Dotted curve shows the PSF of 3.2 mJy source, and dashed
  (solid) curves plot the SZ profiles with (without) the possible
  point source contribution, respectively, using the best-fit
  parameters in the X-ray observation. }
\label{fig:350}
\end{center}
\vspace*{0.2cm}
\begin{center}
  \leavevmode\epsfysize=10cm \epsfbox{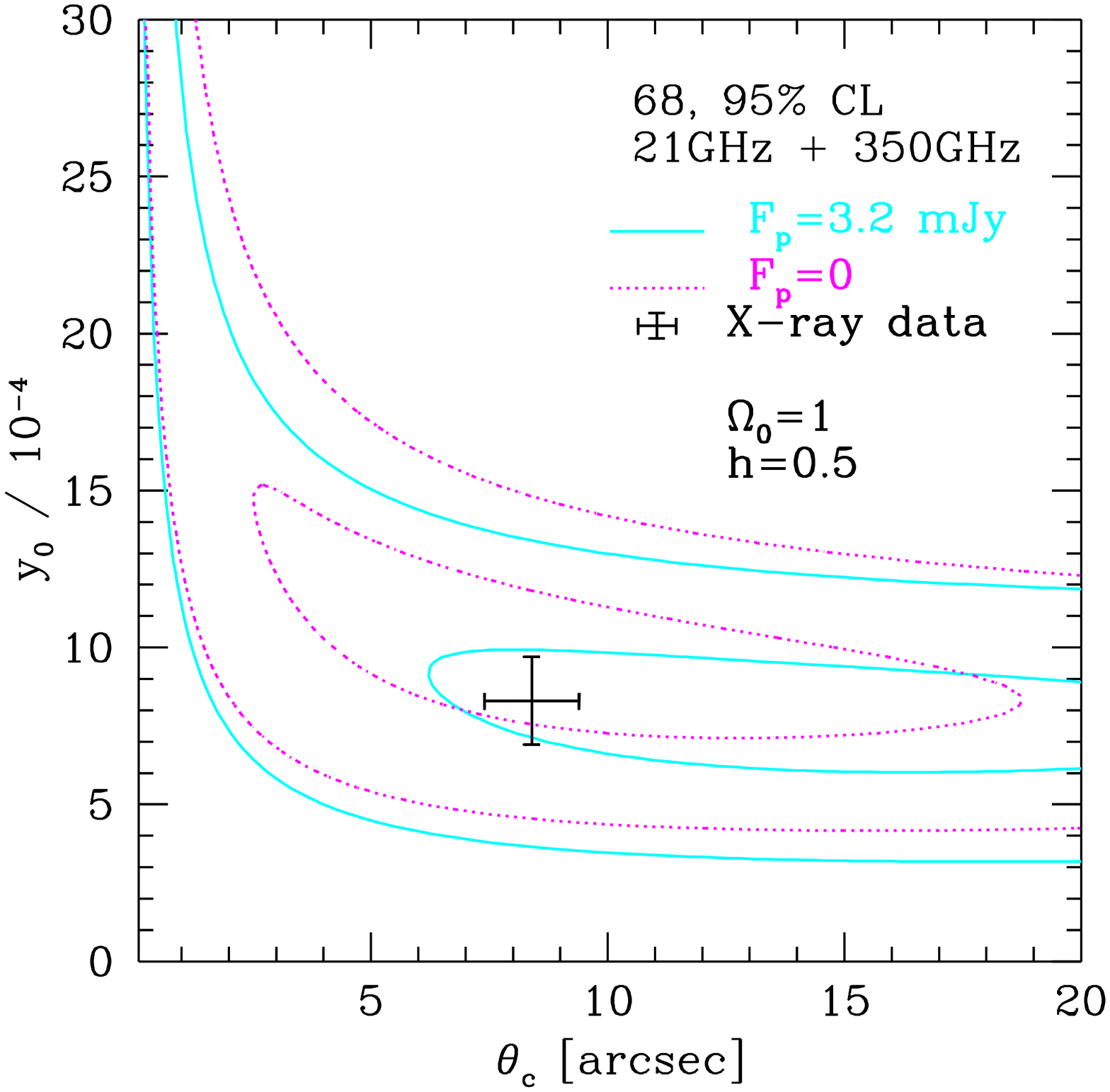}
\caption{\baselineskip=14pt
  Confidence contours on the $y$-parameter and the core radius
  $\theta_{\rm c}$ from combined data analysis at 21 and 350 GHz.  The
  cross indicates the parameters determined from X-ray observations by
  ROSAT and ASCA satellites.}
\label{fig:ythetac}
\end{center}
\end{figure}

The observation of the cluster at 350 GHz with SCUBA was performed on
May 30 and 31, 1998.  The raw data were processed with REMSKY in SURF
package to remove spatially correlated sky-noise.  Since REMSKY
systematically changes the base-level of the map up to the sky-noise
level, the resulting zero-level is uncertain depending on the sky
condition. We estimated 1$\sigma$ error of our base-level or DC
offset, $I_{\rm DC}$, to be as large as $2.9$ mJy/beam; the sky
condition during our observation was not so good (the zenith optical
depth at 350 GHz ranged around $\tau_{350} = 0.46 - 0.60$).  Also we
extracted 7 spurious contamination sources above a threshold of
3$\sigma$ using SExtractor package.  The radially averaged profile of
the cluster at 350 GHz {\it after subtracting the sources and the
  uncertain DC level} is plotted in Figure \ref{fig:350}. Also plotted
in the figure are the SZ intensity profiles at 350 GHz predicted from
the latest X-ray observation with the point source of $F_{\rm p}=0$
and 3.2 mJy, as well as the point spread function of the 3.2 mJy
source. Since normal radio sources often show a steep decline of flux
at higher (submillimeter), we regard the value, 3.2 mJy, extrapolated
from equation (\ref{eq:pointflux}) as the upper limit of the flux.
Again it is clear that the extended feature of the observed intensity
profile cannot be described by the point spread function alone.

\clearpage

To express the result in a statistical and quantitative manner, we
carried out a $\chi^2$ fit to the SZ profiles at 21 and 350 GHz
simultaneously on the basis of the $\beta$-model with taking a central
$y$-parameter, $y_{\rm c}$, and the angular core radius, $\theta_{\rm
  c}$, as free parameters. The results are summarized in Figure
\ref{fig:ythetac}. The agreement with the X-ray observation is
remarkable and reassuring, and we conclude that the present data
provide the strongest and most convincing case for the detection of
the submm SZ signal from the cluster.  Detailed analysis taking
account of various astrophysical effects and the cosmological
implications will be presented elsewhere (Komatsu et al.  1999a,b).

\section{DISCUSSION AND CONCLUSIONS}

It is widely recognized now that many cosmological models are more or
less successful in reproducing the structure at redshift $z\sim0$ {\it
  by construction}. This is because the models have still several
degrees of freedom or {\it cosmological parameters} which can be
appropriately {\it adjusted} to the observations at $z\sim0$
($\Omega_0$, $\sigma_8$, $h$, $\lambda_0$, $b(r,z)$).  There are
several ways to break such cosmological degeneracy. For instance,
combining the {\it COBE} 4 year data with the cluster Log $N$ -- Log
$S$ observation, we showed that low-density CDM models with
$(\Omega_0,\lambda_0,h,\sigma_8) = (0.3,0.7,0.7,1)$ and $(0.45, 0,
0.7, 0.8)$ are viable models, while the Einstein -- de Sitter model is
strongly disfavored even if equation (\ref{eq:sigma8logns}) can be
satisfied by adjusting $\sigma_8$ accordingly.

Along with such consideration, we proposed a cluster survey based on
the SZ effect. As a first step toward the goal, we conducted the
search for the SZ effect towards the cluster RXJ1347-1145 in
multi-bands. We detected the extended signal which are in good
agreement with the predicted SZ profiles, especially at 21 and 350
GHz.  In particular, our results provide the strongest and most
convincing case for the detection of the SZ temperature increment in
the submm band in which the profile of a cluster of galaxies is
resolved for the first time.

\vspace*{1cm}

We thank Iain Coulson, Nario Kuno and Satoki Matsushita for kind
assistance during our observing runs at JCMT, Nobeyama 45-m and
Nobeyama Millimeter Array, respectively. We also thank Nick Tothill
for providing the calibration data for our observation, and Tim
Jenness, John Richer, Remo Tilanus and Goeran Sandell for many
fruitful comments and suggestions on data analysis via the SCUBADR
mailing list.  We are also grateful to John Carlstrom for information
of the flux of the central source at 28.5 GHz. The travel of E. K. to
Hawaii was supported in part by Satio Hayakawa Foundation in the
Astronomical Society of Japan. T. K. gratefully acknowledges support
from a JSPS (Japan Society for the Promotion of Science) fellowship.
This research was supported in part by the Grants-in-Aid for the
Center-of-Excellence (COE) Research of the Ministry of Education,
Science, Sports and Culture of Japan to RESCEU (No.07CE2002).

\clearpage

{\large\bf REFERENCES}

\vspace{-5mm}
\begin{itemize}
\setlength{\itemindent}{-8mm}
\setlength{\itemsep}{-1mm}

\item[] Andreani, P., {\it et al.} {\it Astrophys.J.}, {\bf 459}, L49 (1996)
\item[] Bryan, G. L., Norman, M. L. {\it Astrophys.J.}, {\bf 495}, 80
  (1998)
\item[] Bunn, E. F., White, M. {\it Astrophys.J.}, {\bf 480}, 6 (1997)
\item[] Cagnoni, I., Della Ceca, R., Maccacaro, T. {\it Astrophys.J.},
  {\bf 493}, 54 (1998)
\item[] Condon, J. J. {\it et al.} {\it Astron. J.}, {\bf 115}, 1693
  (1998)
\item[] Ebeling, H., Edge, A. C., Fabian, A.C., Allen, S. W.,
  Crawford, C. S. {\it Astrophys. J.}, {\bf 479}, L101 (1997)
\item[] Ebeling H., {\it et al.} {\it Mon.Not.R.Astron.Soc.}, submitted
  (1998)
\item[] Eckart, A., Hummel, C. A., Witzel, A. {\it
    Mon. Not. R. Astron. Soc.}, {\bf 239}, 381 (1989).
\item[] David, L. P., Slyz, A., Jones, C., Forman, W., Vrtilek,
  S. D. {\it Astrophys. J.}, {\bf 412}, 479 (1993)
\item[] Kitayama, T., Sasaki, S., Suto, Y. {\it
    Publ. Astron. Soc. Japan}, {\bf 50}, 1 (1998)
\item[] Kitayama, T., Suto, Y. {\it Mon.Not.R.Astron.Soc.}, {\bf 280},
  638 (1996a)
\item[] Kitayama, T., Suto, Y., {\it Astrophys.J.}, {\bf 469}, 480
  (1996b).
\item[] Kitayama, T., Suto, Y., {\it Astrophys.J.}, {\bf 490}, 557
  (1997).
\item[] Komatsu, E., Kitayama, T., Suto, Y., Hattori, M., Kawabe, R.,
   Matsuo, H., Schindler, S., Yoshikawa, K., 
  {\it Astrophys.J.}, in press (1999a)
\item[] Komatsu, E., {\it et al.}, in preparation  (1999b)
\item[] Lacey, C. G., Cole, S. {\it Mon.Not.R.Astron.Soc.}, {\bf 262},
  627 (1993)
\item[] Lamarre, J. M., {\it et al.} {\it Astrophys. J.} {\bf 507}, L5
  (1998)
\item[] Mushotzky, R. F., Scharf, C. A. {\it Astrophys.J.}, {\bf 482},
  L13 (1997)
\item[] Piccinotti, G., Mushotzky, R. F., Boldt, E. A., Holt, S. S.,
  Marshall, F. E., Serlemitsos, P. J., Shafer, R. A. {\it
    Astrophys.J.}, {\bf 253}, 485 (1982)
\item[] Press, W. H., Schechter, P. {\it Astrophys.J.}, {\bf 187}, 425
  (1974)
\item[] Rephaeli, Y. {\it ARA\&A} {\bf 33}, 541 (1995) 
\item[] Rephaeli, Y., Yankovitch, D. {\it Astrophys.J.}, {\bf 481},
  L55 (1997)
\item[] Rosati, P., Della Ceca, R., Burg R., Norman, C., \& Giacconi,
  R. {\it Astrophys.J.}, 445, L11 (1995)
\item[] Rosati, P., Della Ceca, R., Norman, C., Giacconi, R. {\it
    Astrophys. J.}, {\bf 492}, L21 (1998)
\item[] Sasaki, S. {\it Publ. Astron. Soc. Japan}, 46, 427 (1994)
\item[]Schindler, S., Hattori, M., Neumann, D. M.  B\"ohringer, H.
  {\it Astron. Astrophys.}, {\bf 317}, 646 (1997) 
\item[] Sunyaev R.A., Zel'dovich Ya.B. {\it Commts.\ Astrophys.\ Space
    Phys.}, {\bf 4}, 173 (1972)
\item[] Ueda, Y., Takahashi, T., Inoue, H., Tsuru, T., Sakano, M.,
  Ishisaki, Y., Ogasaka, Y., Makishima, K., Yamada, T., Ohta, K.,
  Akiyama, M.  {\it Nature}, {\bf 391}, 866 (1998)
\end{itemize}

\end{document}